\RequirePackage[2020-02-02]{latexrelease}

\documentclass{iau}

\usepackage{amsmath}
\usepackage{graphicx}
\usepackage{multirow}

\begin{document}

\lefttitle{Cambridge Author}
\righttitle{The Galactic Extinction Horizon}

\jnlPage{1}{7}
\jnlDoiYr{2023}
\doival{10.1017/xxxxx}

\aopheadtitle{Proceedings IAU Symposium}
\editors{Fatemeh Tabatabaei,  Beatriz Barbuy,  \& Yuan-Sen Ting, eds.}

\title{The Galactic Extinction Horizon with Present and Future Surveys}

\author{Dante Minniti}
\affiliation{Institute of Astrophysics, Faculty of Exact Sciences, Universidad Andres Bello, Santiago, Chile}
\affiliation{Vatican Observatory, V00120 Vatican City State, Italy}
\affiliation{Departamento de Física, Univ. Federal de Santa Catarina, Trindade 88040-900, Florianópolis, SC, Brazil}
\begin{abstract}
We have made a lot of progress in the study of the MW. In spite of this, much of our Galaxy remains unknown, and amazing breakthroughs await to be made in the exploration of the far side of the Galaxy. Focussing on the Galactic extinction horizon problem with current surveys like the Two Micron All-Sky Survey (2MASS) and the Vista Variables in the Via Lactea Survey (VVV) and its extension VVVX, the extinction horizon is a fundamental difficulty, and it is my intention here to reveal how profound is our ignorance, and also to try to suggest ways for improvement with future near-IR Galactic surveys. 
\end{abstract}


\maketitle

\section{Introduction}

In the twentieth century there has been much progress on the study of the MW. But do not think for a minute that we know a lot. As a matter of fact we know very little about our galaxy, and much of it remains unexplored. A lot of observational evidence has been gathered in the Solar neighborhood and in the halo. This evidence allows to study the structure, dynamics, formation and evolution of our galaxy. But at low latitudes our picture is sketchy beyond a few kiloparsecs, as most of the MW is still without reach, and therefore our ignorance still swamps our knowledge. Outstanding regions that must be unveiled are the Galactic bulge, the nuclear region, and the far side of the disk.

How far can we see in the presence of extinction? In the Milky Way disk this is what we call the extinction horizon. The Gaia mission for example has been a revolution because of the accurate photometric and astrometric measurements of billions of field stars. In the Galactic plane, however, these optical measurements are limited to a distance of a few kiloparsecs because of extinction. 
The IR surveys alleviated the situation, starting with the Two Micron All-Sky Survey (2MASS) that showed us a beautiful all-sky picture of the Milky Way as an edge-on spiral galaxy (Skrutskie et al. 2006), and the Galactic Legacy Infrared MidPlane Survey Extraordinaire (GLIMPSE) survey (Benjamin et al. 2003).
In the Galactic plane, however, the 2MASS and GLIMPSE visions were limited ($Ks < 14-15$ mag), and were complemented with the deeper  ($Ks < 17-18$ mag) and higher resolution UKIDSS Galactic Plane Survey (Lucas et al. 2008), and the VISTA Variables in the Via Lactea survey (VVV) and its recent extension VVVX survey (Minniti et al. 2010, Saito et al. 2023). 

\section{The VVV + VVVX surveys}

The 4m diameter VISTA telescope at ESO Paranal Observatory is IR optimized, and its instrument VIRCAM covers a wide field of view (1.5 sq.deg.).
This telescope was used for the VVV and VVVX surveys, that were awarded $>400$ nights of observation. 
The VVV observations started in year 2010 and were completed in 2016. 
The VVVX observations were extended in area and time, starting in year 2017 and  completed in early 2023.
The overall coverage of the surveys is approximately 1700 sq.deg. (about 4\% of the sky).
They contain multi-epoch measurements (about 30 to 300 epochs) for $>2\times 10^{9}$ sources, 
about $10^{7}$ of which are variable  in the near-IR.
The surveys database contains photometry, light curves and proper motions, that are becoming publicly available {\footnote{vvvsurvey.org
archive.eso.org}}.
As  advanced products, we have released 
the VVV Near-IR Variability Catalog (VIVA) (Ferreira Lopes et al. 2020),
the near-IR Catalog of known variable stars (Herpich et al. 2021), and the
variable star periods and classification across the MW disk and bulge (VIVACE)  (Molnar et al. 2022).
The VVVX observations enable a variety of variability studies about
RR Lyrae (Minniti et al. 2016, 2017a, Majaess et al. 2018, Contreras Ramos et al. 2018),
Classical Cepheids (Minniti et al. 2020),
Type 2 Cepheids (Braga et al. 2018),
Miras and Long Period Variables (LPVs) (Molnar et al. 2020, Sanders et al. 2022, Nikzat et al. 2022),
Eclipsing Binaries (Gramajo et al. 2020),
YSOs (Lucas et al. 2017, Contreras Pena et al. 2017, Guo et al. 2022),
Microlensing (Navarro et al. 2017, 2018, 2020abc), Husseiniova et al. 2021,
Novae and CVs (Saito et al. 2013), and also of unknown objects that we named
WITs (What Is This?)(Minniti et al. 2019, Lucas et al. 2020, Saito et al. 2021, Smith et al. 2021).
In addition, the total baseline for proper motions in our survey is now $>12$ yr.
Complementing with data from  2MASS, the total baseline available for bright sources ($K_s < 15$) in common in these regions is $>20$ yr.

Using the near-IR observations in the Galactic plane region we can use different distance indicators to reach out farther, and to extend the extinction horizon. Some examples of these tracers are:

— {\bf Open clusters:} they are moderately bright and can have good or acceptable distances, but their space density is low and therefore they are poor tracers of the extinction horizon. 

— {\bf Globular clusters:} that are also very bright and can be seen to large distances, but they are rare ad therefore also poor to determine the Galactic extinction horizon.

— {\bf RR Lyrae:} they are excellent distance indicators, and also very numerous across the plane and bulge. They are acceptable tracers to probe the extinction horizon, although their blue colors make them fainter in the near-IR.

— {\bf Type 2 Cepheids:} : they are also excellent distance indicators, and more luminous than RR Lyrae, although also more rare. They can aid the RR Lyrae as tracers of the extinction horizon.

— {\bf Classical Cepheids:} also excellent distance indicators and much more luminous than RR Lyrae, they are relatively rare, and therefore less preferable to map the extinction horizon.

— {\bf Miras/LPVs:} they are very luminous distance indicators, and also numerous across the Galaxy, allowing to extend the extinction horizon to large distances.

— {\bf Red Clump (RC) giants:} they are very numerous, and also relatively bright in the near-IR. Even though they are not as good distance indicators as some of the variable stars, their space density makes them the ideal tracers to map the extinction horizon.

— {\bf Galaxies:} background galaxies are  very numerous, although they are faint in general. Also their spatial distribution is non-uniform, but being located beyond our Galaxy, their mere presence is an indication of clear lines of sight throughout our Galaxy.

A word of caution is that even for the best distance indicators, inside the Milky Way plane the distances are very hard to determine, and certainly dependent on a number of observational selection effects. Extinction provides a very peculiar limitation, and without discussing the complexity of the ISM and extinction laws, my emphasys here would be on pointing out how deep is our ignorance on the stars and structure of the MW from an observers point of view.

\section{The extinction horizon}

How far can we see in the presence of extinction?
Of course how far can we see depends on what we are interested in. 
We can use different distance indicators mentioned above to trace the extinction horizon along the Galactic plane.
Using these tracers, for the different surveys one can make maps of the limiting distance vs Galactic longitude, restricting the observations to $|b| < 2$ deg.

In the optical the limitation is severe within the Galactic plane.
For example, Gaia photometry reaches $G=21.0$ mag. This means that we could see Solar type stars out to a distance of about 10 kpc in the absence of extinction. However, taking into account the recent high-quality optical extinction maps based on Gaia (Lallement et al. 2022, Vergely et al. 2022), within the Galactic plane this distance is halved due to extinction.
Of course in the near-IR the situation is much improved. Considering the recent IR extinction maps for the inner bulge (Gonzalez et al. 2013, Schultheis et al. 2014, Surot et al. 2020) and disk extinction maps (Soto et al. 2019), the VVV survey can reach Solar type stars out to about 10 kpc, fully covering the Galactic bulge, and also RC giants all across the Galactic plane.

The Galactic plane is densely covered with  RC giants, and as distance indicators they are very convenient to explore the Milky Way.
When exploring the extinction horizon using RC Giants from the VVV Survey we observe some dips in the spatial density distribution that may be due to the spiral arm tangents.
We also observe a void beyond the Galactic centre region, which is due to the combined effect of extinction and crowding. 
That is a very special place that needs to be explored in detail.

Note here the importance of the extinction windows. The classical optical windows like Baade’s window have been a revolution for our understanding of the stellar populations in the Galactic bulge (e.g., Stanek 1996). In recent years, the near-IR extinction windows allow us to pierce through the whole Galaxy, tracing structures in the far side (Minniti et al. 2018, Saito et al. 2020, Kammers et al. 2022). 
Nevertheless, pencil beam surveys are limited, and deep surveys covering a much larger area are necessary.

In particular, the presence of galaxies is an excellent indication for windows of low extinction. Indeed, the increased  density of galaxies traces well the low extinction regions at the lowest Galactic latitudes (Baravalle et al. 2021). However, because galaxies are clustered, it is difficult to establish a direct correlation between number of galaxies and amount of extinction.

\section{How many stars are there in the MW?}

How many stars are there in the MW? We still do not know the answer to this basic question. A very crude estimate may be obtained assuming that the total mass of the Galaxy is approximately $10^{12}$ (Watkins et al. 2019), yielding a total number of stars of  a few $10^{11}$ if we account for the presence of dark matter. Yet our largest surveys have measured a few $10^{9}$ stars in total. For example, 

-- there are  $\sim 2\times 10^{9}$ stars (point sources) in total with measured photometry in the VVV+VVVX surveys of the Galactic plane (Smith et a. 2018, 2022 in preparation, Alonso Garcia et al. 2019, 2022 in preparation); 

-- there are  $\sim 3.3 \times 10^{9}$ stars (point sources) in total with measured photometry in the Dark Energy Camera Plane survey  (DECAPS)  (Schlafly et al. 2028, Saydjari et al. 2023);

-- there are  $\sim 2.9 \times 10^{9}$ stars (point sources) in total with measured photometry in the PANSTARRS all-sky survey (Magnier et al. 2020); and

--  there are $\sim 2 \times 10^{9}$ stars measured by Gaia over the whole sky (Brown 2021, Gaia Collaboration 2022), among others.

This means that we have only measured about 1\% of the MW stars! Put it in another way, we have not seen 99\% of the MW stars, and do not know anything about them (nor about their planetary systems). We know the direction of the sky where most of these the stars should be located, towards the Galactic plane and bulge, but we have not been able to observe them. 
Why? Due to two major problems towards these regions: extinction and crowding.  These limitations define
the extinction horizon in the Galactic plane, beyond which we are blind.
What do we need to push this extinction horizon? We need higher resolution, and deeper photometry in the near-IR throughout the whole plane.

\section{How many star clusters are there in the MW?}

Star clusters are also good probes within the Milky Way. However, the total number of star clusters in the MW is still unknown. Of course this number is very difficult to estimate not only because of the extinction horizon problem, but also because of field contamination. In addition, it is risky to extrapolate the local findings to larger distances because we do not really know if the MW is a homogeneous and symmetric structure, and there are disruption processes that act to selectively destroy some kinds of clusters in different regions.  But let us assume that a good estimate is probably a few $10^{4}$ MW star clusters in total. Of these, there are about $10^{4}$ candidate star clusters known to date in the MW (e.g., Bica et al. 2019). 
Or could there be more than $10^{5}$ MW star clusters in total? These crude estimates help to illustrate how much we do not know.

The majority of these are open clusters and associations, with only a couple hundred globular clusters known. The globular clusters are important old survivors, and their numbers are incomplete, especially in the inner bulge regions, where we probably  know only about $2/3$ of the total globular cluster population (Bica et al. 2016, Minniti et al. 2017b, 2021). 

There has also been a notable progress in cluster searches due to the near-IR surveys like 2MASS and VVV (e.g., Froebrich et al. 2007, Borissova et al. 2011, 2014, Chene et al. 2012, Barba et al. 2015), and to the Gaia mission (e.g., Cantat-Gaudin et al. 2018, 2020, Castro-Ginard et al. 2022). Still,  we have only identified a fraction of the MW star clusters. Our ignorance is deep, and it is sobering to realize that there are some clusters that we will never be able to study/identify. Many of them would be forever hidden behind dozens of magnitude of extinction, while for others the membership determination would be impossible. In particular, determination of cluster members is very tricky, as we need 3D information (accurate positions, distances, radial velocities and PMs) in addition to accurate chemical compositions. It is only possible to estimate the physical parameters such as masses, ages, luminosities, metallicities, sizes, orbits, etc., for those clusters that have the most complete set of observations.

\section{Some pathways to make progress}

Within the current paradigm that the MW formed through a series of accretion events that continue until present times, and given the diversity of galaxies (observed or simulated),
it is very dangerous to assume that the well known portion of the Milky way is representative of the whole galaxy.
It is also risky to adopt symmetry arguments that have not been demonstrated observationally.

Remember that we currently see a tiny fraction of the stars, inside the extinction horizon (which varies depending on the tracer and also on wavelength). 
Even though the current panorama is gloomy, there are clear ways to move forward. What do we need? 
It seems to me that the most efficient way to make significant progress 
is to increase this observed stellar fraction up to $>50\%$ as a first priority. 
We need large surveys of the MW plane with higher resolution and deeper photometry.

The survey capabilities of the Vera Rubin Telescope (a.k.a. LSST) would be an improvement, although paradoxically, it would be easier to study the distant outer halo than the far side of the MW disk. For example, we will probably learn more about the Magellanic Clouds as galaxies than our own.

The Nancy Grace Roman Space Telescope (a.k.a. WFIRST) would be the best tool, offering much higher resolution and deeper photometry in the near-IR.
In particular, we need the baseline for variability and proper motion studies.
There is one Roman telescope concept survey proposed that can make significant progress.
Paladini et al. (2022) have proposed the alactic Roman Infrared Plane Survey (GRIPS) survey with the Roman space telescope, that will map the Galactic plane to
unprecedented depth, allowing us to measure a significant fraction of the MW stars, as well as to make a 3-D near IR extinction map of the Galaxy.

There is additional hope: the Roman plus Rubin telescopes combination will multiply the number of stars that we can measure (a.k.a. R2-D2, Bentz et al 2022).
Note that there will be now a Ks-band filter available for the Roman space telescope (Stauffer et al. 2018).
This is very important, providing not only depth in high extinction regions, but also high resolution for accurate PM measurements.

As mentioned, the exquisite Gaia DR3 parallaxes have allowed to map the local volume, 
In the works there is also Gaia IR, a proposed ESA mission that would be the IR Gaia equivalent, that would measure parallaxes and PMs throughout the MW (Hobbs et al. 2016). However, a big leap would be obtained going to even longer wavelengths into the near-IR, because the Ks-band allows us to extend farther the extinction horizon, and we needed a Ks-band filter for the Roman telescope (Stauffer et al. 2018). This suggestion has been implemented, and now it is possible to reach much more heavily obscured regions deep in the Galaxy.

Will we ever see all the MW stars? 
It makes no sense to pretend to observe a reddened WD or BD at D=20 kpc on the far side of the Galactic plane, that would indeed be prohibitive  even with the future missions. But we cannot deny that it would be of outmost importance to have a full census of all the RR Lyrae, and all the Cepheids for example. And as another example, to map also distant Solar type stars across the whole disk of the Galaxy. This is not a project so crazy to do if we are going to look for civilizations akin to ours with ever more powerful telescopes like the Square Kilometer Array (SKA).

Unfortunately there are no planned telescopes/missions that will see all the MW stars, so this will probably not happen for many decades.
Why do we care? Because we need to make progress in several fronts, e.g.:

-- The RRLyrae census in the Galactic center is incomplete (Minniti et al. 2016, Navarro et al. 2022).

-- Microlensing events are useful probes of the mass distribution in the Galaxy. However, their usefulness is distance-limited in ways that we do not know, because we are still missing many distant microlensing events with sources located in the far side of the Galaxy
(Navarro et al. 2020b).

-- Hypervelocity stars (HVS) are useful probes of the interactions with the black hole at the Galactic center (Brown 2015). However, in order to accurately estimate what is the production rate of HVS from the supermassive black hole we need to sample the HVS in their creation site (e.g., Luna et al. 2019).

-- If part of the plane is hidden, we may miss the next Galactic supernova explosion. Or even if we detect the event, we may miss the identification of the progenitor  (e.g., Adams et al. 2013).

-- Speculating farther, there is much effort concentrated in searching for planets around Solar type-stars. 
But if we do not detect most of the Galactic Solar type stars, we may miss the 
the parent stars of technological civilizations like ours (or even more advanced civilizations) (e.g., Ivanov et al. 2020).

-- In spite of all previous efforts, the Galactic centre region has not been mapped thoroughly at high spatial resolution with existing ground and space-based telescopes. Even more, we have no clear observations of our Galaxy within the region beyond the Galactic centre. While crowding is a severe limitation for such observations, the higher resolution of the Roman  telescope will alleviate the  incompleteness  due to severe crowding.

-- Could we detect a dwarf galaxy located towards the far side of the Galactic plane? Probably not, even though this is not a far fetched possibility. There is mounting evidence that some globular clusters are the remaining nuclei of disrupted dwarf galaxies, that long ago were accreted by the MW. In fact, even bright globular clusters may have been missed at low latitudes due to high extinction. The recent discovery of the massive globular cluster FSR1758 (Cantat-Gaudin et al. 2018, Barba et al. 2019), associated with one of the largest past Galactic mergers called the Sequoia event, serves as a warning that bright objects (globular clusters, dwarf galaxies) may still be discovered in the far side of the Galaxy.

-- We worry about the star formation history of the Milky Way, without having a complete current star formation map of our Galaxy to start with. In addition, we are still searching for massive young clusters (e.g., Chene et al. 2012). But for example, we still do not know if there are just a couple of Carina-like star forming regions in the Galaxy, or a couple dozen.

We need a deep high resolution near-IR survey of the Galactic plane with the Roman telescope not only to measure a larger fraction of all the MW stars, but also to answer some basic questions:
How are the different stellar (and planet) populations and the gas and dust distributed within the MW?
Where are most of the heavy remnants (neutron stars, black holes) located?
Where will the next significant Galactic gravitational wave event occur?, etc.

\section{Conclusions}

For now the main conclusion (that we can also call unavoidable truth) remains that the MW is an immense entity about which we know very little.
Not only there are important details missing, but also huge observational gaps. For example, only $\sim 1$\% of the MW stars have been measured to date. In particular, the far side of the Galaxy remains largely unknown, and can be probed nowadays only through small windows of low extinction.

The future deep photometry and high resolution imaging capabilities of the Rubin Telescope and the Roman Space Telescope will mitigate the severe incompleteness problems due to crowding and extinction towards the Galactic plane, allowing to increase dramatically the sources with adequately measured positions and photometric baselines.
In particular, it is very important for future studies of the MW to have a wide near-IR survey of the Galactic plane and bulge from space, that would also feed targets for the JWST and the ELTs.
\\

\noindent 
{\bf Acknowledgments:} 
D.M. gratefully acknowledges support by the ANID BASAL projects ACE210002 and FB210003, and by Fondecyt Regular Project No. 1220724.

\end{document}